\documentclass[aps,prd,nofootinbib]{revtex4}
\usepackage{graphicx}
\usepackage{amsfonts}
\usepackage{amsmath}
\usepackage{amssymb}

\begin{document}

\title{On the Kalb-Ramond modified Lorentz violating hairy black holes and Thorne's hoop conjecture}
\author{K.K. Nandi}
\email{kamalnandi1952@rediffmail.com}
\affiliation{Zel'dovich International Center for Astrophysics, Bashkir State Pedagogical University, 3A, October Revolution Street, Ufa 450008, RB, Russia}
\affiliation{High Energy Cosmic Ray Research Center, University of North Bengal, Siliguri 734013, WB, Bharat}
\author{R.N. Izmailov}
\email{izmailov.ramil@gmail.com}
\affiliation{Zel'dovich International Center for Astrophysics, Bashkir State Pedagogical University, 3A, October Revolution Street, Ufa 450008, RB, Russia}
\author{R.Kh. Karimov}
\email{karimov_ramis_92@mail.ru}
\affiliation{Zel'dovich International Center for Astrophysics, Bashkir State Pedagogical University, 3A, October Revolution Street, Ufa 450008, RB, Russia}
\author{A.A. Potapov}
\email{zicastro.ufa@gmail.com}
\affiliation{Ufa University of Science \& Technology, Sterlitamak Campus, Lenin Avenue, 49, Sterlitamak 453103, RB, Russia}

\date{31 October 2023}

\begin{abstract}
Recently, a class of static spherically symmetric power law corrected Lorentz violating (LV) Schwarzschild black holes in the Kalb-Ramond model have been derived and studied in the specific range of LV parameters ($0<\lambda \leq 2,\Upsilon \geq 0$) that correspond to energy condition preserving ($\rho >0$) source. On the other hand, there exist well known black holes that do not preserve the energy conditions. In this paper, we shall therefore relax energy conditions and numerically explore the horizon patterns of the enlarged class of LSMA black holes. Four generic types of LV corrected black holes emerge, which interestingly include the analogue of the \textit{braneworld} black hole ($\rho <0$) lending to $\Upsilon$ a new interpretation of "tidal charge" known as an imprint from the $5d$ bulk in the Randall-Sundrum scenario. We shall then show that Thorne's hoop conjecture, $\mathcal{H} \leq 1$, where $\mathcal{H}$ is the Hod function, consistently holds for three types and their generalizations. However, intriguingly, it turns out that, for the remaining type (viz., Schwarzschild-de Sitter and its generalizations), the hoop conjecture does \textit{not} hold. It is also shown that braneworld tidal charge black holes increases the LV correction to planetary perihelion advance in contrast to the decrease due to ordinary black holes thereby providing a qualitative distinction between them.
\end{abstract}

\maketitle


\section{Introduction}\label{sec1}
In previous papers \cite{Nandi:2020, Nandi:2022}, it was shown how the Hod function $\mathcal{H}$ could be used to determine the status of Thorne's hoop conjecture \cite{Thorne:1972} for the existence of black hole horizons in static spherically symmetric solutions belonging to general relativity or modified theories. There have been numerous useful works on this famous conjecture in various contexts -- it is impossible to list them all here (nevertheless, see e.g., \cite{Hod:2015, Peng:2019, Peng:2021, Hod:2019, Senovilla:2008, Gibbons:1997, Malec:1991, Redmount:1983, Chiba:1994, Abrahams:1992, Nakamura:1988, Liberati:2022, Ida:1998, Flanagan:1991, Murchadha:2010, Anza:2017}).%
\footnote{%
Hod \cite{Hod:2020a} showed that the conjecture, though it holds for static black holes, does not hold for the spinning ones. He proposed an entirely new geometric conjecture for spinning black holes \cite{Hod:2020b} that has also been supported by various spinning solutions across theories of gravity \cite{Nandi:2021}.}

The methodology developed in \cite{Nandi:2020, Nandi:2022} can be applied to a new class of power law modified Lorentz violating (LV) static spherically symmetric hairy Schwarzschild black holes in the Kalb-Ramond model recently derived by Lessa, Silva, Maluf and Almeida (hereinafter LSMA) \cite{Lessa:2020}, which is the object of this paper. LSMA analyzed the effects of the LV parameters $\lambda$ and  $\Upsilon$ mainly within the range ($0<\lambda \leq 2,\Upsilon \geq 0$) in order that the corresponding matter source satisfied the energy conditions. However, for arbitrary choices of these parameters beyond the preceeding range, the source of the black hole could very well be violating the energy conditions. We might call them "exoticized" Schwarzschild black hole spacetimes to distinguish them from the ordinary ones having spacetimes satisfying the energy conditions ($\rho >0$) or traversable wormholes that necessarily require exotic matter ($\rho <0$) as their sources.\footnote{%
Exotic matter is defined as the matter violating the Null Energy Condition (NEC) defined by $\rho +p_{r}\geq 0$ \cite{Visser:1995} It takes only a uniformly moving observer to see the Weak Energy Condition (WEC) violation ($\rho <0$) as NEC violation, i.e., $\rho +p_{r}<0$ \cite{Morris:1988}.} Exotic black hole spacetimes do exist in the literature. A well known example is the Schwarzschild-Anti de Sitter (SAdS) black hole [$\lambda =-1,p=+1,\rho <0$, see Eq.(12) below] having a wormhole-like topology since Ricci scalar $\mathbf{R}<0$. Another example is the \textit{braneworld tidal charge} black hole ($\lambda =1,p=-1$, $\rho <0$, $\mathbf{R}=0$), in which the tidal charge is not electric but is a correction to the Schwarzschild black hole, just as the LV corrections are, except that the tidal corrections appear as an imprint from the $5d$ bulk onto the $3d$ brane in the Randall-Sundrum scenario \cite{Randall:1999}. The LSMA class of black holes include the above two examples as special cases. One can recover the Schwarzschild black hole when the LV corrections are set to zero.

Since here we are going to deal with Lorentz violating black holes, it makes sense to specify the physical motivation for considering such a violation. In fact, the motivation comes from the deficiencies of the standard model of particle physics. Although the model agrees very well with experiment, many important questions still remain unanswered, e.g., it does not account for the existence of dark matter and dark energy, which make up most of the energy density of the universe but are not composed of any known particles covered by the standard model. The minimal $SU(3)\times SU(2)\times U(1)$ standard model is believed to be the low-energy limit of an extended theory that can also provide a plausible theory of quantum gravity. To obtain a general extension of the minimal standard model (see e.g., \cite{Colladay:1998}) yielding the low-energy limit, the idea of spontaneous Lorentz violation (LV) has been used. The violation in the gravity sector can be caused by the presence of nonzero tensor expectation values in the vacuum. In the present case, it's the background Kalb-Ramond field providing the nonzero vacuum expectation value and contributing parameters or hairs ($\lambda, \Upsilon$) to the LSMA solution modifying the Schwarzschild black hole.

Turning now to the investigation of hoop conjecture for such black holes, the mass-circumference ratio describing it can be generically rephrased in terms of a Hod function $\mathcal{H}$ based on the Misner-Sharp quasi-local mass $M(r)$ that includes all sorts of mass-energy within a radius $r$ in any static spherically symmetric spacetime (this mass is a direct result of Einstein's field equations, see \cite{Lynden-Bell:2007, Nandi:2009, Misner:1964}). But $M(r)$ does not lead to asymptotic ADM mass $M_{\infty}$ in the non-flat spacetimes in which horizons may nevertheless exist. In such cases, it is more appropriate to use Hayward quasi-local energy $E(r)$ in the extended global coordinate chart and define the apparent horizons as trapped surfaces (see, for details, the pivotal paper \cite{Hayward:1996}). However, the Hod function, dedicated to defining only the hoop conjecture, involves the quasi-local mass $M(r)$ enclosed within a finite hoop radius in standard coordinates. In these coordinates, black hole thermodynamics have also been studied in the literature in the asymptotically non-flat single horizon spacetimes like the SAdS (see Hawking \&\ Page \cite{Hawking:1983}) or the double horizon Schwarzschild-de Sitter (SdS) discussed by Roy Choudhury and Padmanabhan \cite{Roy:2007}. The latter work concludes that there is no single invariant global Hawking temperature associated with SdS spacetime and that the situation is not as straighforward as with a single horizon spacetime. We shall show that the hoop conjecture does not hold in SdS, which might not come as a surprise given the problematic thermodynamic picture of SdS.

The purpose of the present paper is to numerically study, following the methodology developed in \cite{Nandi:2020, Nandi:2022}, the generic pattern of the appearance of horizons and the status of the hoop conjecture in the LSMA black holes for different values of the LV parameters not necessarily restricted to the specific range ($0<\lambda \leq 2,\Upsilon \geq 0$). We shall also briefly discuss the LV contribution to the precession of planetary orbits in the braneworld type black hole. Hereafter, we add the qualification "type" to the LV corrected black holes to distinguish them from the well known general relativistic ones despite their formal resemblence because the Kalb-Ramond stress properties sourcing the LV black holes are different from those of general relativity. However, for arbitrary values of $\lambda$ except $1,-1$, the resulting solutions do not even resemble those of general relativity thereby lending novelty to the present work since we shall choose  $\lambda$ arbitrarily.

The plan of the paper is as follows: Sec.2 introduces Thorne's hoop conjecture in terms of Hod function with a glimpse into its global definition. In Sec.3, we briefly sketch the Kalb-Ramond field equations and its solution
following \cite{Lessa:2020} and in Sec.4, the conjecture is implemented for numerical analyses of LSMA black hole horizons. Sec.5 presents the results of the analyses. Sec.6 summarizes the work with two speculative remarks. We take units such that $G=1,c=1$.

\section{Thorne's hoop conjecture, Hod function, trapped surfaces}\label{sec2}
The hoop conjecture, proposed by Thorne \cite{Thorne:1972}, states that an imploding object forms a black hole when, and only when, a circular hoop with a specific critical circumference could be placed around the object and rotated about its diameter in all directions. Symbolically, horizons exist when and only when a mass $M$ gets compacted into a region whose circumference $C$ in every direction is bounded by $C\leq 4\pi M$, giving the circumference-mass ratio as
\begin{equation}
\frac{C}{4\pi M}\leq 1
\end{equation}%
An alternative statement of the hoop conjecture is that any circumference $C$ on the horizon is bounded by $C\leq 4\pi M$ \cite{Tomimatsu:2005}. However, the mass in the above ratio was not clearly specified that gave rise to counter-examples \cite{Bonnor:1983, deLeon:1987} contradicting the conjecture. If the mass $M$ is considered to be the asymptotic ADM mass $M_{\infty}$ of a \textit{horizonless} charged spacetime, it might be so arranged as to yield an inequality $\frac{C}{4\pi M_{\infty }}<1$ implying the occurrence of a horizon, while the spacetime is horizonless by construction. It was proven in \cite{Bonnor:1983} that the hoop conjecture can be contradicted in charged fluid sphere spacetimes.

This contradiction was first resolved by Hod \cite{Hod:2018}, who interpreted that the mass in the denominator in (1) should be the mass $M(r\leq R)$ contained within a hoop of minimum circumference $C$ $(\equiv 2\pi R)$ and not the total asymptotic mass $M_{\infty}$, i.e., the ratio (1) be replaced by%
\begin{equation}
\mathcal{H=}\frac{C}{4\pi M(r\leq R)}\leq 1\Rightarrow \text{ existence of horizon,}
\end{equation}%
where we had called $\mathcal{H}$ the Hod function in \cite{Nandi:2020}, $R$ the circumferential areal radius of the smallest hoop engulfing the mass in all azimuthal directions. For the charged spacetime with charge $Q$, the concerned mass should be, according to Hod \cite{Hod:2018},%
\begin{equation}
M(r\leq R) = M_{\infty} - \frac{Q^{2}}{2R}.
\end{equation}%
This redefined mass resolves the contradiction posed in \cite{Bonnor:1983}. Generically, the mass $M$ in (2) for any static spacetime with stress tensor $T_{\mu\nu}$ should be defined by \cite{Nandi:2020, Lynden-Bell:2007}
\begin{equation}
M(r\leq R)=M_{\infty }-\int_{R}^{\infty }T_{0}^{0}4\pi r^{2}dr.
\end{equation}%
The hoop conjecture can be simply expressed as a circumference-mass ratio as 
\begin{equation}
\mathcal{H\leq }1\Rightarrow \frac{R}{2M(r\leq R)}\mathcal{\leq }1.
\end{equation}%
It was shown in \cite{Nandi:2022} that the definition of physical mass $M(r\leq R)$ within the ball of radius $R$ defined in (3), viz., $M_{\infty }-\frac{Q^{2}}{2R}$ is exactly the same as the Misner-Sharp \textit{geometric} quasi-local mass $M(r\leq R)$ in the Reissner-Nordstr\"{o}m spacetime. For a generic static spherically symmetric spacetime, the Misner-Sharp quasi-local mass $M(r\leq R)$ is defined by \cite{Misner:1964}%
\begin{equation}
M(r\leq R) = \frac{R}{2} \left(1 - \left. g^{rr}\right\vert_{r=R}\right),
\end{equation}
where $g^{rr}$ is the contravariant $rr-$component of the metric tensor of the spacetime. The definition (6) reduces to the ADM mass $M_{\infty}$ at spatial infinity only in the asymptotically flat spacetimes. In the non-flat spacetimes, the $g^{rr}$ diverges at infinity but here we take $M(r\leq R)$ being defined only within a hoop radius $r=R$. Thus, it is convenient to adopt the geometric definition (6) that can be equally used for asymptotically flat or non-flat spacetimes as long as we are focused only on the \textit{finite} hoop radius.

Nevertheless, a glimpse of the global definition of horizon might be informative as to how it could be related to the function $\mathcal{H}$. The LSMA solutions are asymptotically non-flat depending on the nonzero LV parameters and in such cases, it is most appropriate to define an apparent horizon by the global concept of trapped surfaces in extended spacetimes, which requires that the mass $M$ be replaced by the Hayward \textit{energy} $E$. Kodama \cite{Kodama:1980} proved that $E$ is the conserved total energy flux of matter and gravitational field. The extended static spherically symmetric metric is defined, up to a diffeomorphism, by the null coordinates ($\xi ^{+},\xi ^{-}$) as 
\begin{equation}
ds^{2} = r^{2} d\Omega^{2} - 2e^{f} d\xi^{+}d\xi,
\end{equation}%
where $r$ and $f$ are functions of ($\xi ^{+},\xi ^{-}$) and $\Omega$ is the metric on a unit sphere, $r$ is the area radius defined by $4\pi r^{2}$. Define the expansions by $\theta_{\pm} = \frac{2}{r} \partial_{\pm}r$, where $\partial_{\pm}$ are the derivatives along the coordinate directions  $\xi^{\pm}$. The Hayward energy $E(r)$ can be rewritten in terms of these expansions as \cite{Hayward:1996}%
\begin{equation}
E=\frac{r}{2}(1-g^{rr})=\frac{r}{2}+e^{f}r\partial _{+}r\partial _{-}r=\frac{r}{2}+\frac{1}{4}e^{f}r^{3}\theta _{+}\theta _{-}.
\end{equation}

A metric sphere is said to be trapped if $\theta _{+}\theta _{-}<0$, untrapped if $\theta _{+}\theta _{-}>0$ and marginal if $\theta _{+}\theta_{-}=0$. Restated, it means that a metric sphere is trapped if and only if $\frac{r}{2E}<1$, marginal if and only if $\frac{r}{2E}=1$ and untrapped if and only if $\frac{r}{2E}>1$. Future and past trapped spheres occur in black and white holes respectively \cite{Chiba:1994, Hayward:1996, Kodama:1980}. Thus the central quantity is $\frac{r}{2E}$, which controls the existence of trapped spheres and it is the same as in (6) defining $\mathcal{H}$ in standard coordinates but now expressed in terms of global expansions $\theta_{\pm}$.

\section{Kalb-Ramond theory: A brief outline}\label{sec3}
It is assumed that the Lorentz violation (LV) is driven by a self-interacting antisymmetric second rank tensor, $B_{\mu\nu}$, the so-called Kalb-Ramond (KR) field \cite{Kalb:1973}. Assuming that the potential $V$ has a nonzero vacuum expectation value (VEV) $b_{\mu\nu}$, we are interested in modifying the spherically symmetric black holes driven by the Lorentz violating KR VEV. The Kalb-Ramond field non-minimally coupled to gravity yields a hairy black hole that deforms the Schwarzschild event horizon.

Guided by the gravitational sector of the standard model of particles, a self-interacting potential for the Kalb-Ramond field is introduced in the form $V=V(B_{\mu\nu}B^{\mu\nu}\pm b_{\mu\nu}b^{\mu\nu})$ with a non-vanishing VEV $\left\langle B_{\mu\nu}\right\rangle =b_{\mu\nu}$, which defines a background tensor field with the spontaneous breaking of Lorentz symmetry. The LSMA black holes are then derived by assuming that $b_{\mu\nu}$  has a constant norm $b^{2}=b_{\mu\nu}b^{\mu\nu}$.

The action for a self-interacting Kalb-Ramond field non-minimally coupled to gravity has the form \cite{Altschul:2010}
\begin{eqnarray}
S^{\text{KR}} &=& \int\sqrt{-g}d^{4}x\left[  \frac{\mathbf{R}}{2}-\frac{1}{12}H_{\lambda\mu\nu}H^{\lambda\mu\nu}-V\left(  B_{\mu\nu}B^{\mu\nu}\pm b_{\mu\nu}b^{\mu\nu}\right)\right. \nonumber \\
&&\left.+\frac{1}{2}\left(  \xi_{2}B^{\lambda\nu}B_{\nu}^{\mu}R_{\lambda\mu}+\xi_{3}B^{\mu\nu}B_{\mu\nu}\mathbf{R}\right)  \right],
\end{eqnarray}
where $H_{\lambda\mu\nu}=\partial\lbrack_{\lambda}B_{\mu\nu}]$ and $\xi_{2}\xi_{3}$ are non-minimal coupling constants with dimension $\left\vert \xi\right\vert \sim\lbrack L]^{2}$. The assumed constancy of $b^{2}$converts the term $\xi_{3}B_{\mu\nu}B^{\mu\nu}\mathbf{R}$ into $\xi_{3}b^{2}\mathbf{R}$ so that it can be absorbed in the pure Ricci term. By varying the metric, one then has the full field equations $G_{\mu\nu}=8\pi GT_{\mu\nu}^{\left(\xi_{2}\right)  }$ (we omit the lengthy expression for $T_{\mu\nu}^{\left(\xi_{2}\right) }$, see \cite{Lessa:2020}). Assuming a metric ansatz of the form
\begin{equation}
d\tau^{2} = -A(r) dt^{2} + B(r) dr^{2} + r^{2}(d\theta^{2} + \sin^{2}{\theta} d\varphi^{2}),
\end{equation}
the field equations yield, for dimensionless $\lambda=\left\vert b\right\vert^{2}\xi_{2}\neq2$, the two equations $A(r)=\frac{1}{B(r)}$ and 
\begin{equation}
\frac{r^{2}\lambda}{2}A^{\prime\prime}+(\lambda+1)rA^{\prime}+A-1=0,
\end{equation}
where prime denotes derivaties with respect to $r_{0}$. The solution is in the power law black hole hole metric \cite{Lessa:2020}
\begin{equation}
A(r)=1-\frac{2M_{\infty}}{r}+\frac{\Upsilon}{r^{\frac{2}{\lambda}}},
\end{equation}
where $r_{S}=2M_{\infty}$ is the Schwarzschild radius, $\Upsilon$ is a constant. The metric has two LV parameters $\lambda$ and $\Upsilon$ (hairs), the latter having dimension $\left\vert \Upsilon\right\vert \sim\lbrack L]^{2/\lambda}$, that control the Lorentz violation effect on the Schwarzschild black hole.

\section{The LSMA black holes: The algorithm for hoop conjecture}\label{sec4}
We define two major characteristics of a classical black hole: it has to have at least one horizon and should satisfy the hoop conjecture since the latter is widely believed to reflect a fundamental aspect of classical general relativity\footnote{%
Despite this belief, its formulation has its limitations, see Hod \cite{Hod:2020a}. Also, of all its various formulations, only some are correct, see Tod \cite{Tod:1992}. We do not go into these discussions in this paper.}. For our convenience, we slightly redefine the LSMA metric function $A(r)$ in \cite{Lessa:2020} by introducing a sign flipper $p$ as follows:%
\begin{eqnarray}
d\tau ^{2} &=&-A_{p}(r)dt^{2}+A_{p}(r)^{-1}dr^{2}+r^{2}(d\theta ^{2}+\sin^{2}\theta d\varphi ^{2}), \\
A_{p}(r) &=&1-\frac{2M_{\infty }}{r}+\frac{p\Upsilon }{r^{\frac{2}{\lambda }}},
\end{eqnarray}%
where $M_{\infty }>0$ is the usual Schwarzschild mass in the absence of LV corrections, $p=\pm 1$ such that we shall maintain $\Upsilon >0$ throughout the analysis, while $\lambda $ can have any sign and value. Note that all of the LSMA black holes have the same Schwarzschild black hole mass $M_{\infty}$. The parameter $\Upsilon$ is a pseudo-electric charge since it does not have the physical attribute of a genuine electric charge for which the stress tensor is traceless, whereas the Kalb-Ramond vacuum expectation value of the pseudo-electric charge is not traceless. For $p=+1$ and $\lambda =1$ belonging to the ranges of LV parameters ($0<\lambda \leq 2,\Upsilon \geq 0$), LSMA \cite{Lessa:2020} studied geometrical features, Hawking temperature and the perihelion advance of planets, the last yielding a tiny value of $\Upsilon_{\lambda =1}$ ($\sim 10^{-3}$ km$^{2}$).

The Ricci scalar for the metric (13) is given by%
\begin{equation}
\mathbf{R} = -\frac{2p\Upsilon (\lambda -1)(\lambda -2)}{\lambda ^{2}}r^{-\frac{2(1+\lambda )}{\lambda}}.
\end{equation}%
The stress components in the orthonormal frame are
\begin{eqnarray}
\rho  &=&\left( \frac{2}{\lambda }-1\right) \frac{p\Upsilon }{r^{\frac{2}{\lambda }+2}}, \\
p_{r} &=&-\rho,  \\
p_{\theta} &=&p_{\varphi }=\frac{\rho }{2}.
\end{eqnarray}%
For convenience, let us define dimensionless quantities $x,s$ as%
\begin{eqnarray}
x &=& \frac{r^{2/\lambda }}{\Upsilon }>0, \\
s &=& \frac{\Upsilon ^{\lambda /2}}{2M_{\infty }}>0,
\end{eqnarray}%
then the metric function becomes 
\begin{equation}
A_{p}(x,s;\lambda) = 1-\frac{1}{sx^{\lambda /2}}+\frac{p}{x}.
\end{equation}%
The algorithm is to first choose some arbitrary value of $\lambda$, say $\lambda_{0}$, and then numerically solve the two simultaneous equations yielding critical values $x_{c},s_{c}$ such that 
\begin{equation}
A_{p}(x_{c},s_{c};\lambda _{0})=0,\left. \partial _{x}A_{p}(x,s;\lambda_{0})\right\vert _{(x_{c},s_{c})}=0.
\end{equation}%
Corresponding to $p=\pm 1$, which give $A_{\pm }$, the two Hod functions, defined in (5) together with (6) and (14) with $g^{rr}=A_{p}(r)$, respectively become
\begin{eqnarray}
\mathcal{H}_{p}(x,s;\lambda _{0}) &=&\frac{sx}{x^{1-\frac{\lambda _{0}}{2}}-ps} \\
\mathcal{H}_{-}(x,s;\lambda _{0}) &=&\frac{sx}{x^{1-\frac{\lambda _{0}}{2}}+s} \\
\mathcal{H}_{+}(x,s;\lambda _{0}) &=&\frac{sx}{x^{1-\frac{\lambda _{0}}{2}}-s}.
\end{eqnarray}%
The next step is to plot the metric function $A_{p}(x,s;\lambda _{0})$. If the dimensionless quantities $x_{c},s_{c}$ are positive and real, then keeping $x=x_{c}$ fixed, the transition between "no horizon" and "horizon" regimes is dimarcated by the plot of $A_{p}(x,s;\lambda _{0})$ of (21) for ranges of $s$ either by \cite{Ayon:1998}: 
\begin{eqnarray}
s &>&s_{c}\Rightarrow \text{no horizon} \\
s &=&s_{c}\Rightarrow \text{coincident horizon} \\
s &<&s_{c}\Rightarrow \text{double horizons}
\end{eqnarray}%
or by its reverse
\begin{eqnarray}
s &<&s_{c}\Rightarrow \text{no horizon} \\
s &=&s_{c}\Rightarrow \text{coincident horizon} \\
s &>&s_{c}\Rightarrow \text{double horizons}.
\end{eqnarray}%

The validity of the conjecture is then verified in the following cases based on the admissible ranges of $s$ yielding horizons obtained from plotting $A_{p}(x,s;\lambda _{0})=0$:

(i) Double horizon: If the inequalities are indicated by the double horizon of $A_{p}(x,s;\lambda _{0})$ according to (26)-(28), then we say that the hoop conjecture will be satisfied in that spacetime if $H_{p}(x,s\leq s_{c};\lambda _{0})\leq 1$ for the ranges of $s$ below the transition point $s=s_{c}$.

(ii) Double horizon: If the reverse inequalities are indicated by the double horizon of $A_{p}(x,s;\lambda _{0})$ according to (29)-(31), then we say that the hoop conjecture will be satisfied in that spacetime if $H_{p}(x,s\geq s_{c};\lambda _{0})\leq 1$ for the ranges of $s$ above the transition point $s=s_{c}$.

(iii) Single horizon: If any of $x_{c},s_{c}$ is either negative or imaginary, they are ruled out as $x>0,s>0$ by definition. In this case, each plot of the metric function shows a single horizon $x=x_{h}$ obtained by solving for any $s$, say, $s_{h}$ the equation $A_{p}(x,s_{h};\lambda _{0})=0,$ and if $\mathcal{H}_{p}(x,s\leq s_{h};\lambda _{0})\leq 1$ holds like (i) above, we say that the hoop conjecture is satisfied. 

\begin{figure}[!ht]
\centering
\includegraphics[scale=0.35]{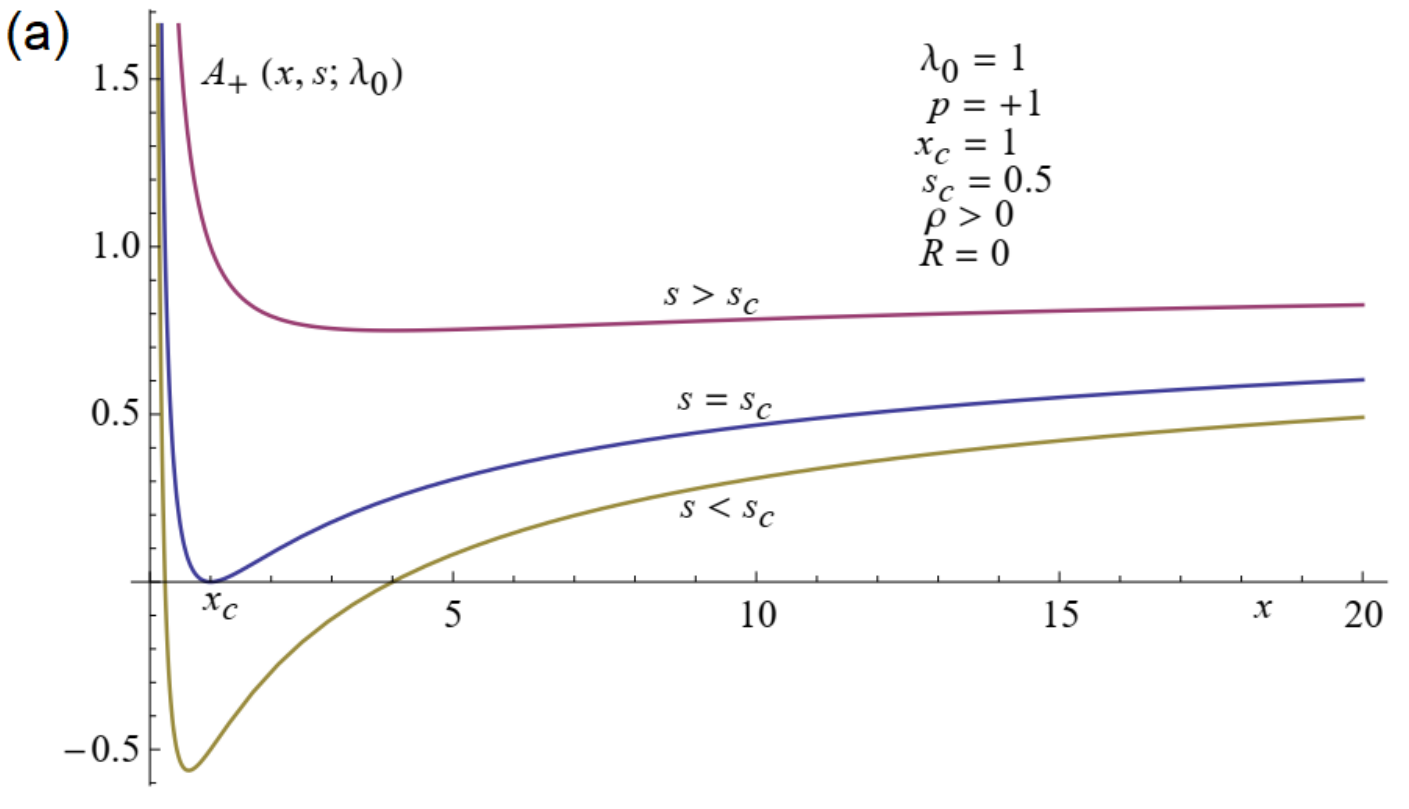} \includegraphics[scale=0.35]{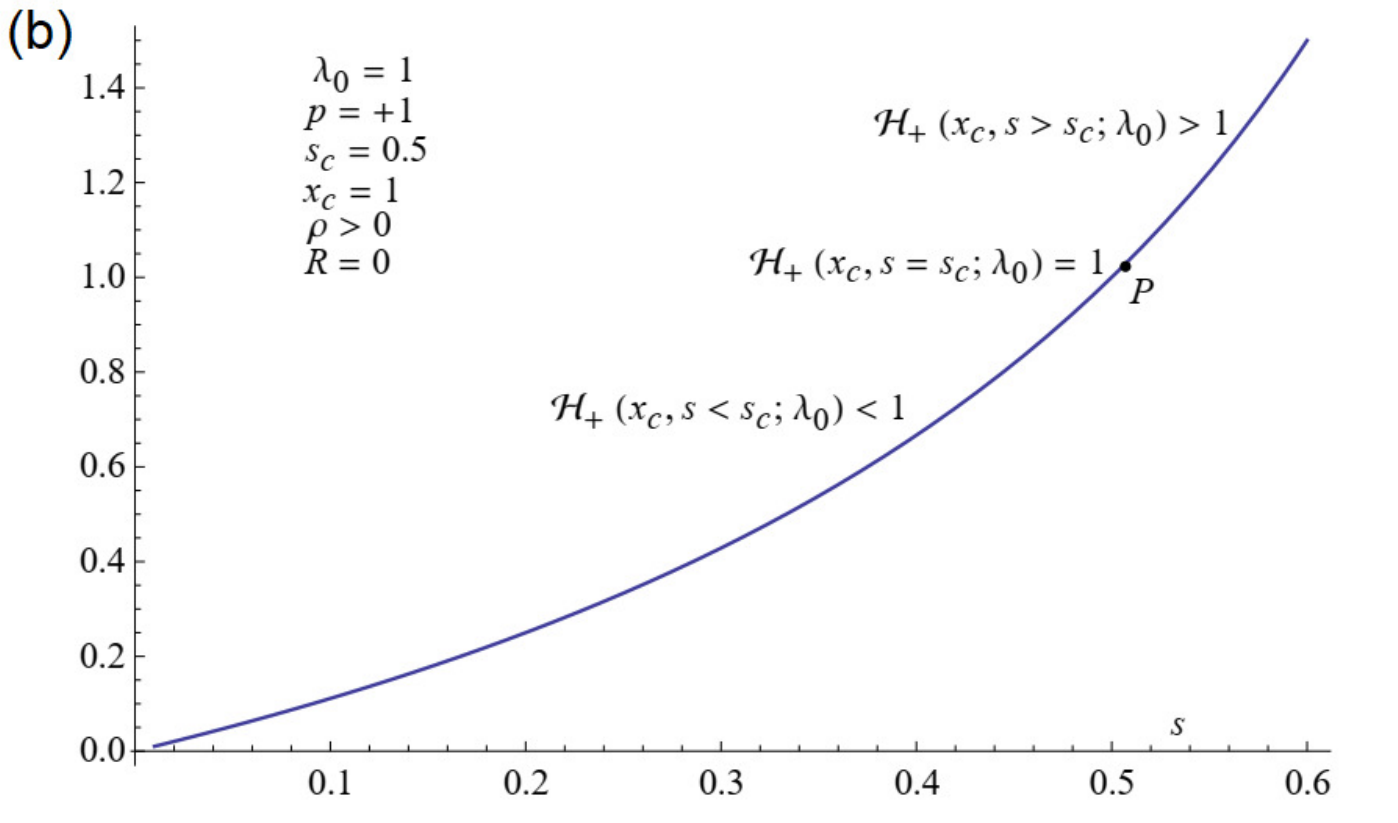}
\caption{(a): The chosen values in the metric function (21) are $\lambda _{0}=1$, $p=+1$ giving an asymptotically flat LV corrected Reissner-Nordstr\"{o}m type metric $A_{+}(x,s;\lambda _{0})$ with double horizons straddling the coincident horizon radius $x=x_{c}$ on either side. Fig.1(b) shows that the hoop conjecture holds: $\mathcal{H}_{+}(x_{c},s\leq s_{c};\lambda _{0})\mathcal{\leq }$ $1$ respecting the ranges dictated by the conditions (26)-(28) for the occurrence of horizons.}\label{fig1}
\end{figure}

\begin{figure}[!ht]
\centering
\includegraphics[scale=0.35]{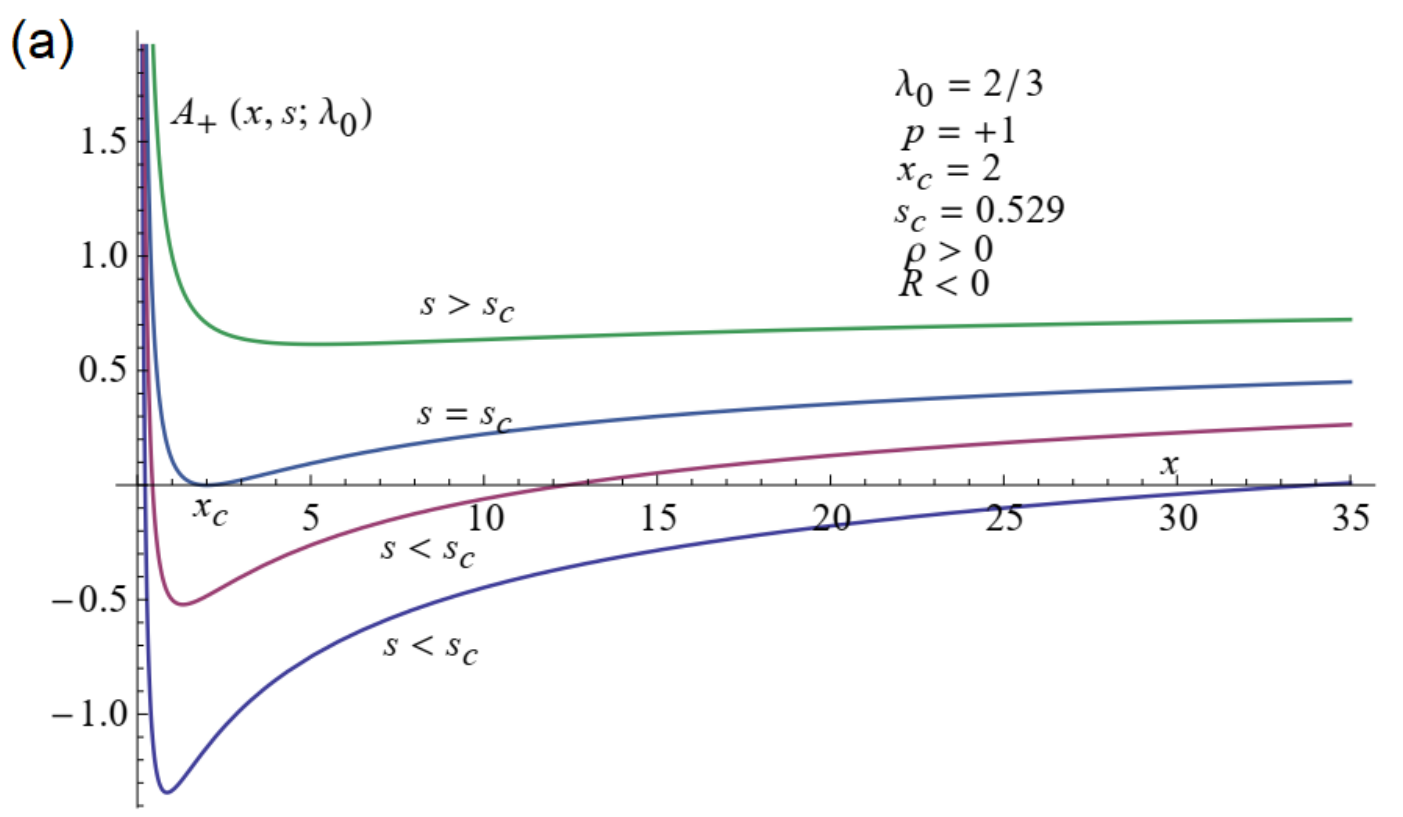} \includegraphics[scale=0.35]{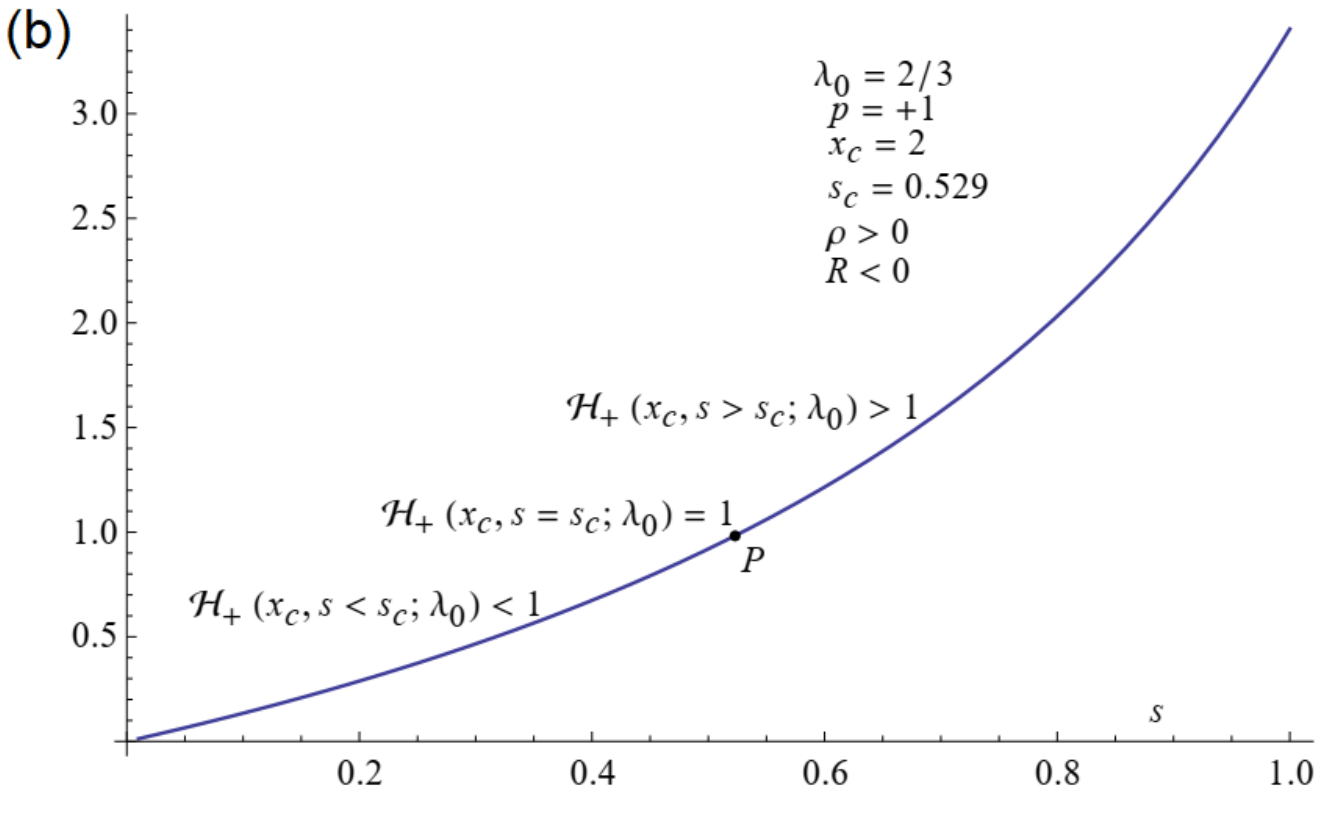}
\caption{(a): The chosen values in the metric function (21) are $\lambda _{0}=2/3$, $p=+1$ giving what we called an asymptotically flat LV corrected generalization of Reissner-Nordstr\"{o}m type metric $A_{+}(x,s;\lambda _{0})$ with double horizons straddling the coincident horizon radius $x=x_{c}$ on either side. Fig.2(b) shows that the hoop conjecture holds: $\mathcal{H}_{+}(x_{c},s\leq s_{c};\lambda _{0})\mathcal{\leq }$ $1$ respecting the ranges dictated by the conditions (26)-(28) for the occurrence of horizons. Any other value $\lambda _{0}>0$ could be chosen at will and it can be verified that the horzon patterns and the hoop conjecture continue to be preserved.}\label{fig2}
\end{figure}

\begin{figure}[!ht]
\centering
\includegraphics[scale=0.35]{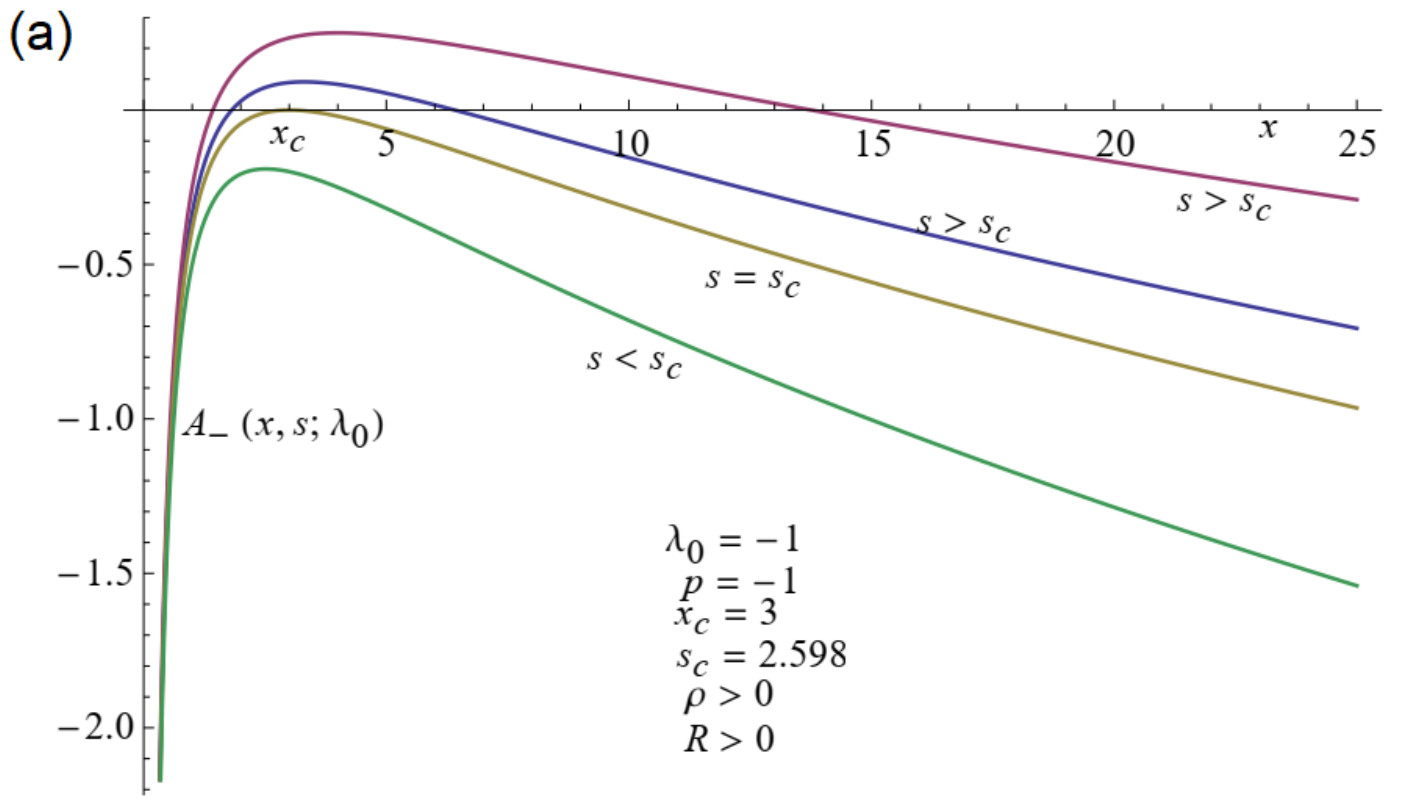} \includegraphics[scale=0.35]{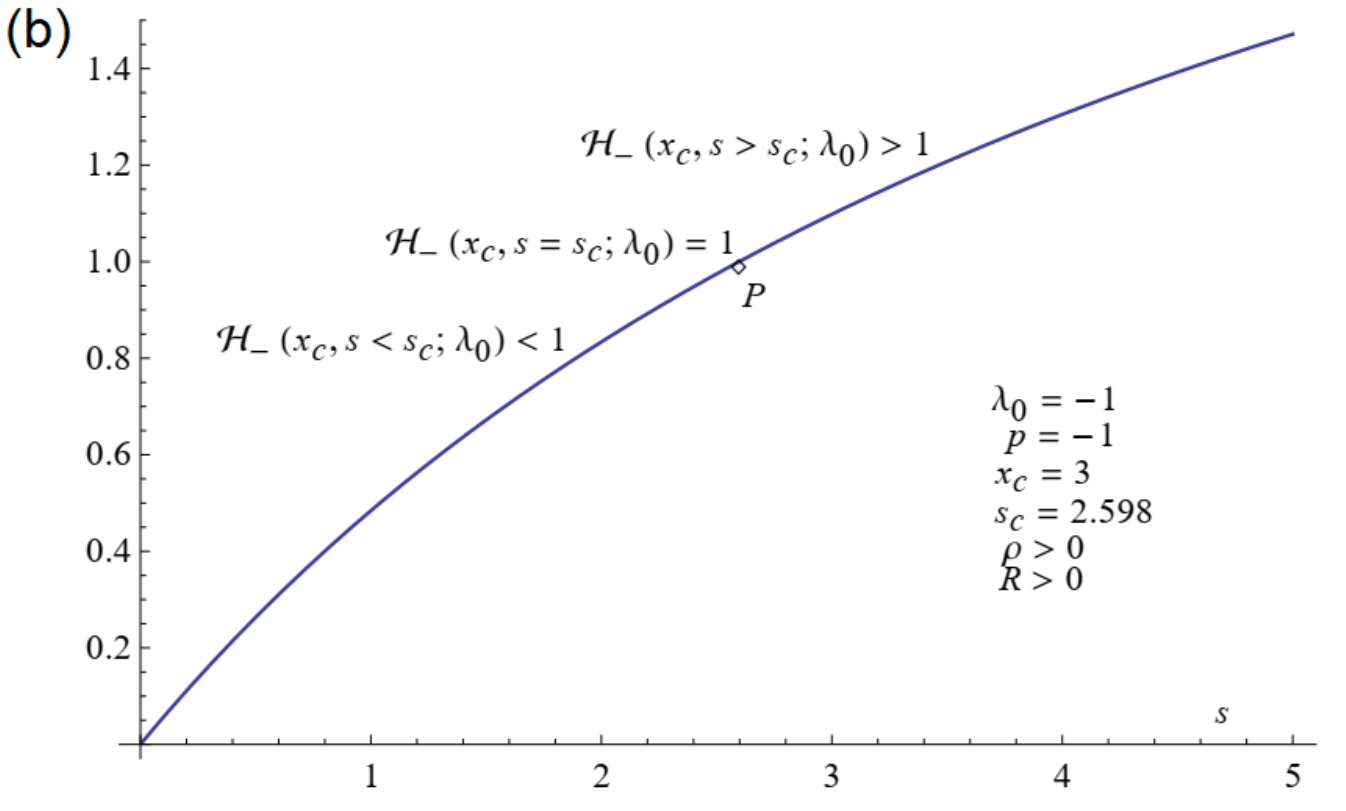}
\caption{(a): The chosen values in the metric function (21) are $\lambda _{0}=-1$, $p=-1$ giving what we called an asymptotically non-flat LV corrected SdS type metric $A_{-}(x,s;\lambda_{0})$ with double horizons straddling the coincident horizon radius $x=x_{c}$ on either side. Fig.3(b) shows that the hoop conjecture does \textit{not} hold in the SdS type spacetime. The reason is that the conjectute $\mathcal{H}_{-}$ $\leq 1$ does not respect the ranges dictated by the conditions (29)-(31) indicating the occurrence of horizons\ in the metric $A_{-}(x,s;\lambda _{0})$ since $\mathcal{H}_{-}(x_{c},s\geq s_{c};\lambda _{0})\mathcal{\geq }$ $1$ contrary to (ii) of Sec.4.}\label{fig3}
\end{figure}

\begin{figure}[!ht]
\centering
\includegraphics[scale=0.35]{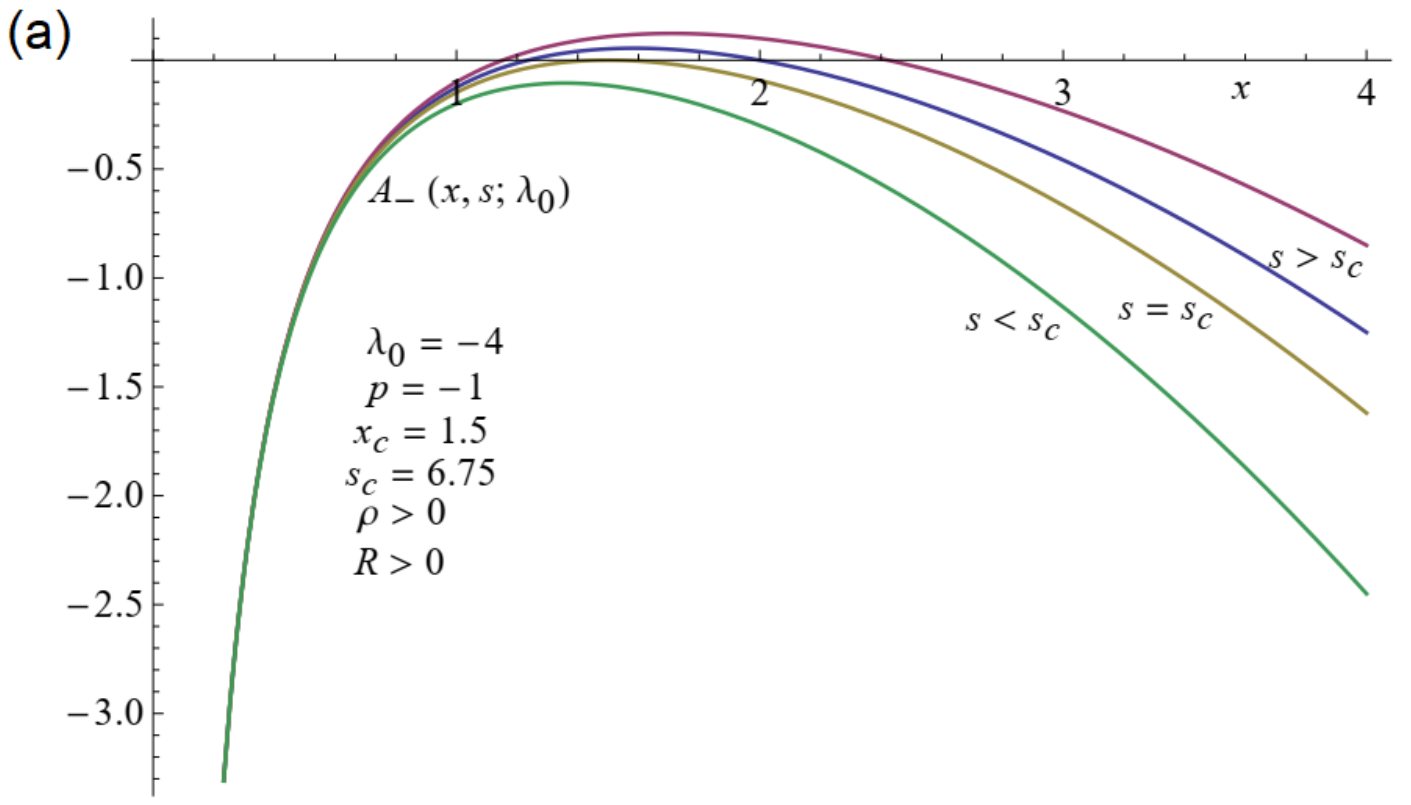} \includegraphics[scale=0.35]{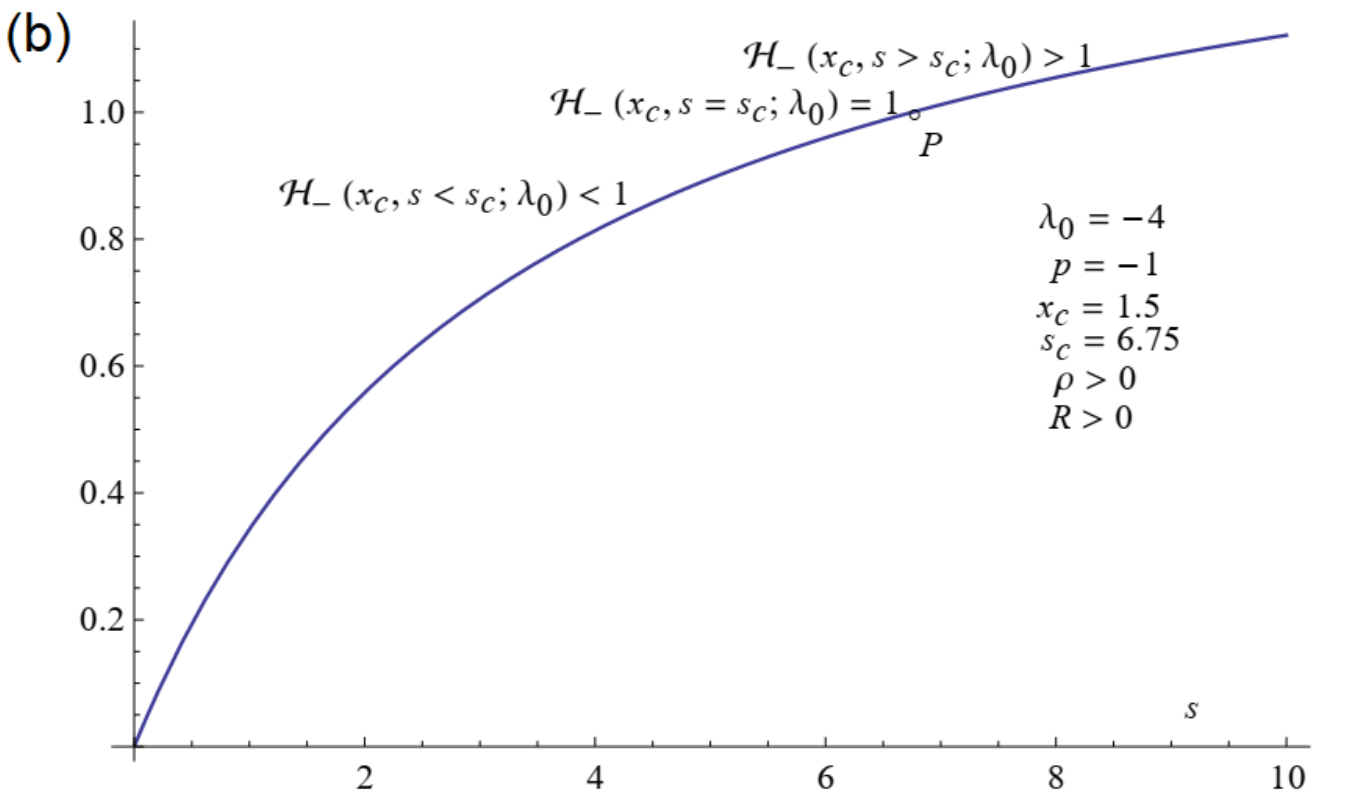}
\caption{(a): The chosen values in the metric function (21) are $\lambda _{0}=-4$, $p=-1$ giving a generalized version of the SdS-type metric. Fig.4(b) shows once again that the hoop conjecture does \textit{not} hold in this generalized SdS type spacetime since the conjecture $\mathcal{H}_{-}(x_{c},s\geq s_{c};\lambda _{0})\leq 1$ does not hold respecting the ranges of $s$ as in (29)-(31) indicating the occurrence of horizons dictated by the plot of the metric $A_{-}(x,s;\lambda _{0})$. Any other value $\lambda _{0}<0$ could be chosen at will and it can be verified that the horizon patterns and the violation of the hoop conjecture continue to be preserved.}\label{fig4}
\end{figure}

\begin{figure}[!ht]
\centering
\includegraphics[scale=0.35]{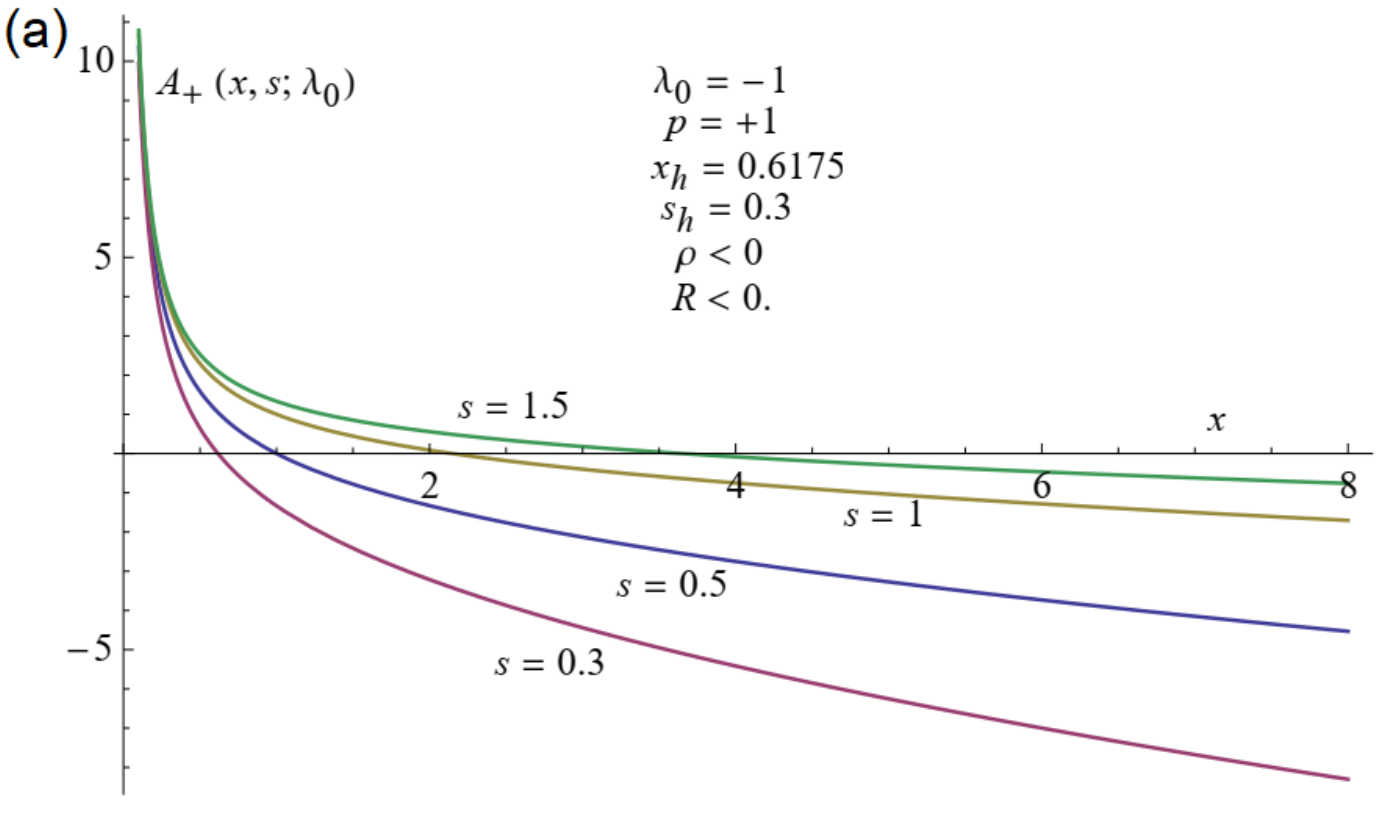} \includegraphics[scale=0.35]{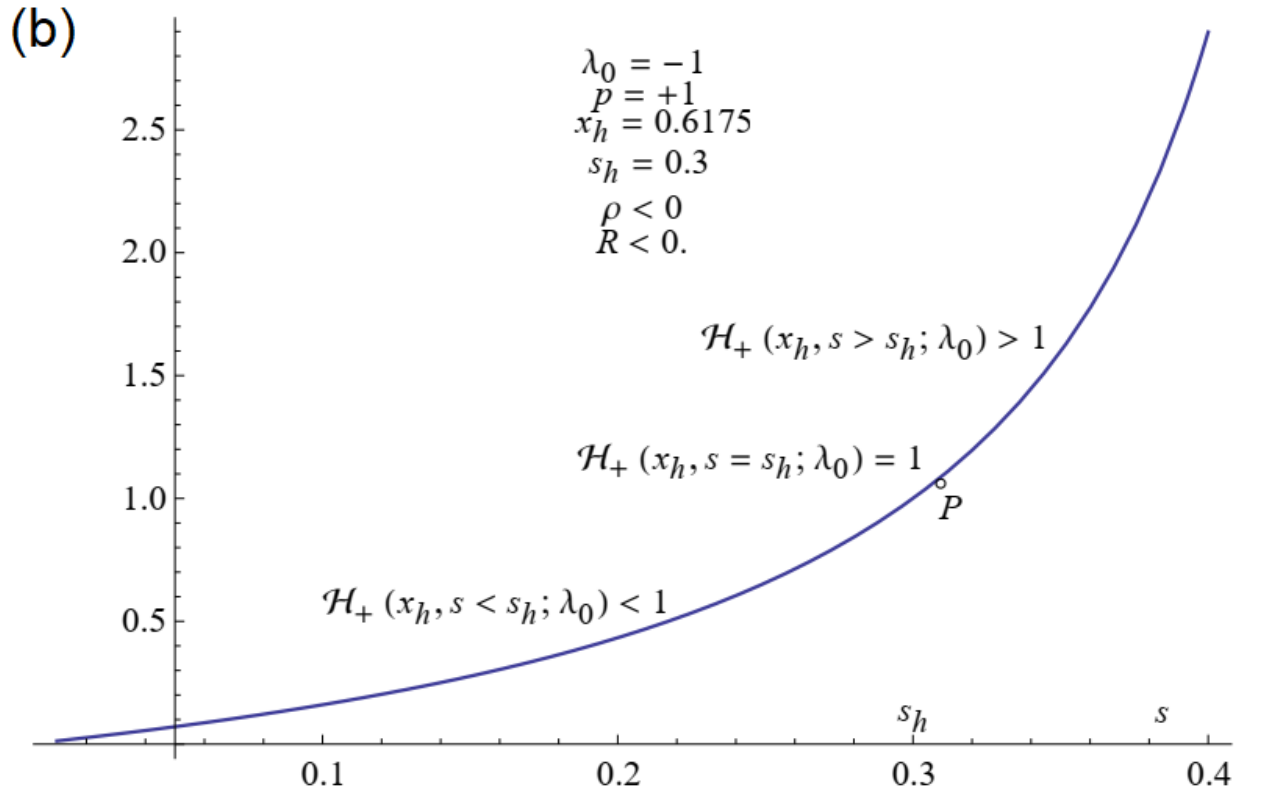}
\caption{(a): The chosen values in the metric function (21) are $\lambda _{0}=-1$, $p=+1$ giving a single horizon SAdS-type metric. Fig.5(b) shows once again that the hoop conjecture holds since $\mathcal{H}_{+}(x_{h},s\leq s_{h};\lambda_{0})\leq 1$ for any horizon radius $x_{h}$ corresponding to a choice $s_{h}$. The point $P$ bifurcates the spacetime into horizon and no-horizon regimes such that in the latter $\mathcal{H}_{+}(x_{h},s>s_{h};\lambda _{0})>1$ holds.}\label{fig5}
\end{figure}

\begin{figure}[!ht]
\centering
\includegraphics[scale=0.35]{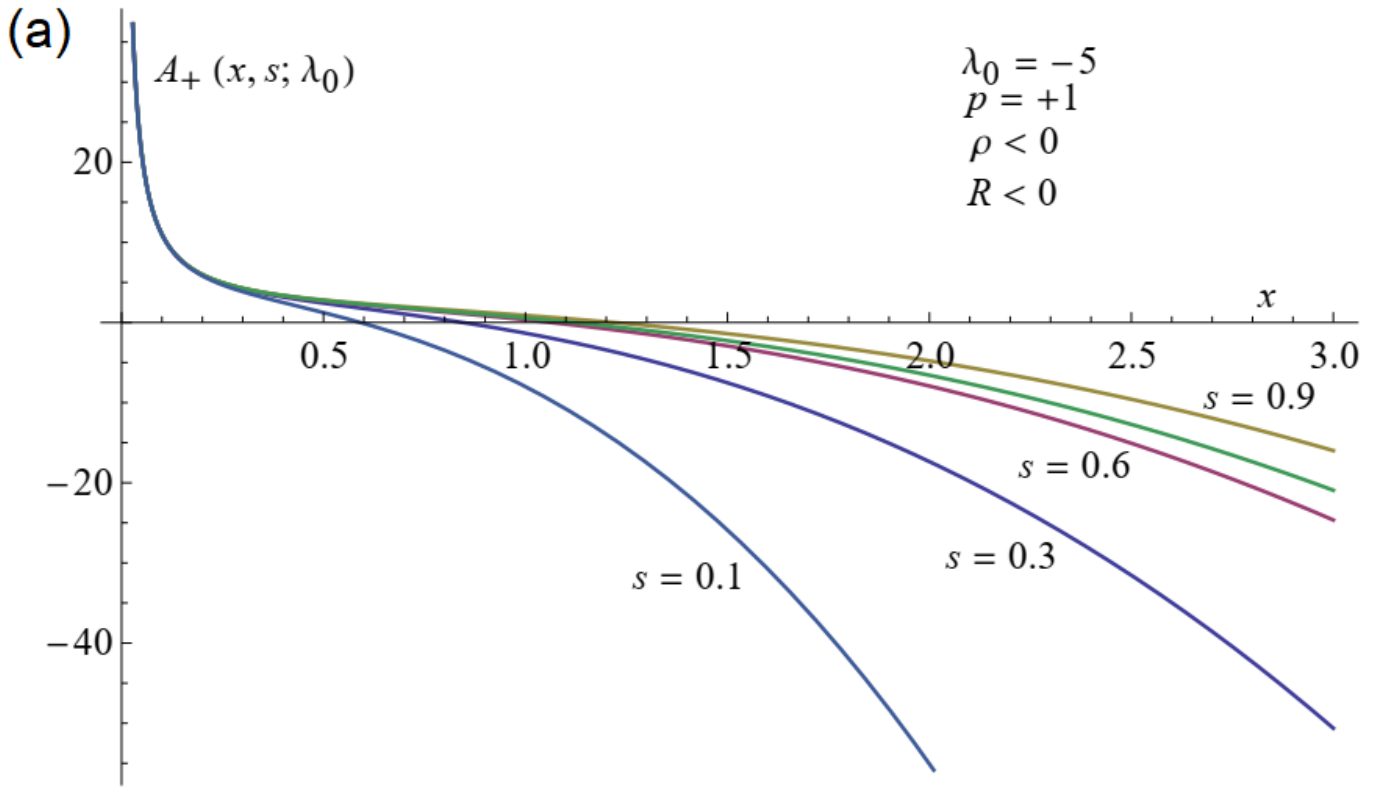} \includegraphics[scale=0.35]{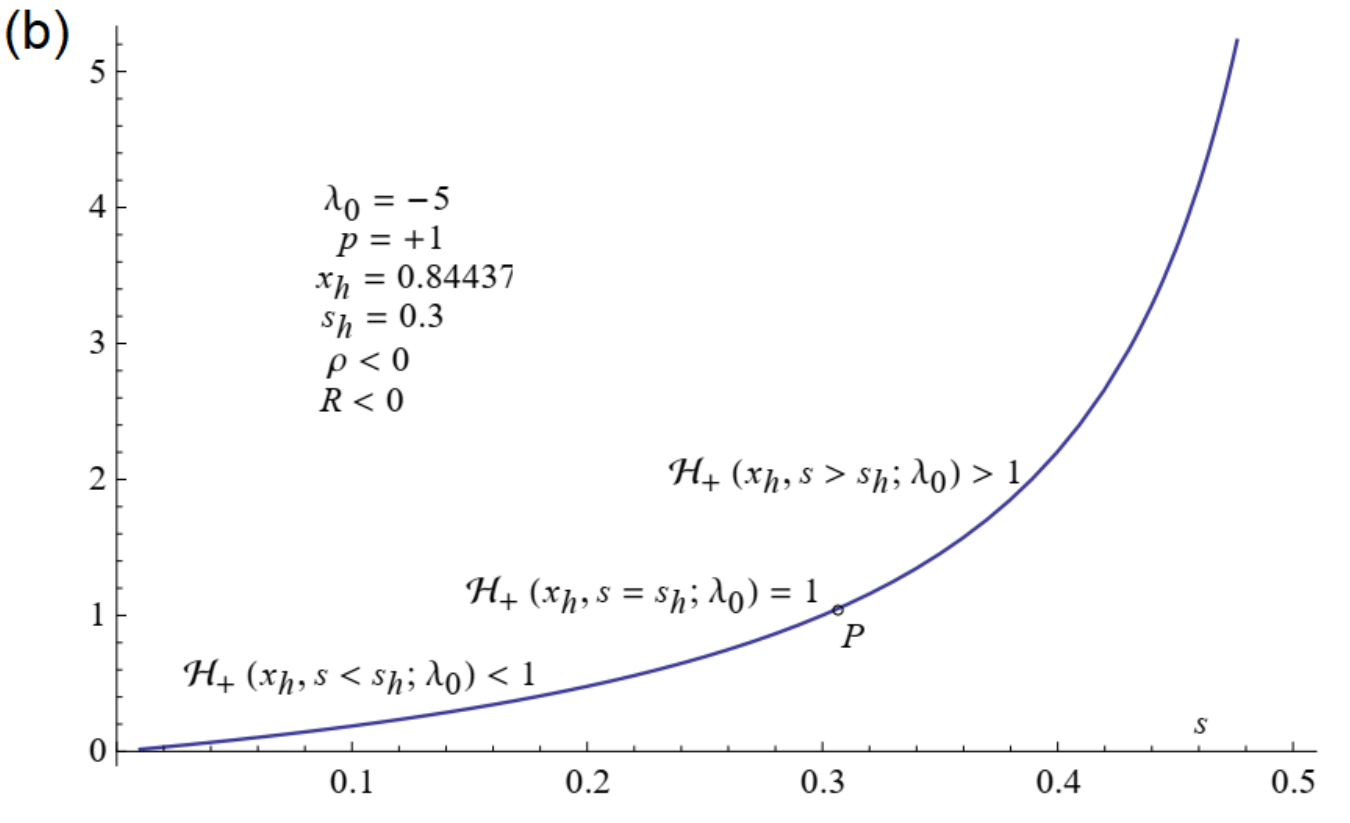}
\caption{(a): The chosen values in the metric function (21) are $\lambda _{0}=-5$, $p=+1$ giving a single horizon generalized SAdS-type metric. Fig.6(b) shows once again that the hoop conjecture holds since $\mathcal{H}_{+}(x_{h},s\leq s_{h};\lambda _{0})\leq 1$ for any horizon radius $x_{h}$ corresponding to a choice $s_{h}$. Any other value $\lambda _{0}<0$ could be chosen at will and it can be verified that the horizon patterns and the hoop conjecture continue to be preserved.}\label{fig6}
\end{figure}

\begin{figure}[!ht]
\centering
\includegraphics[scale=0.70]{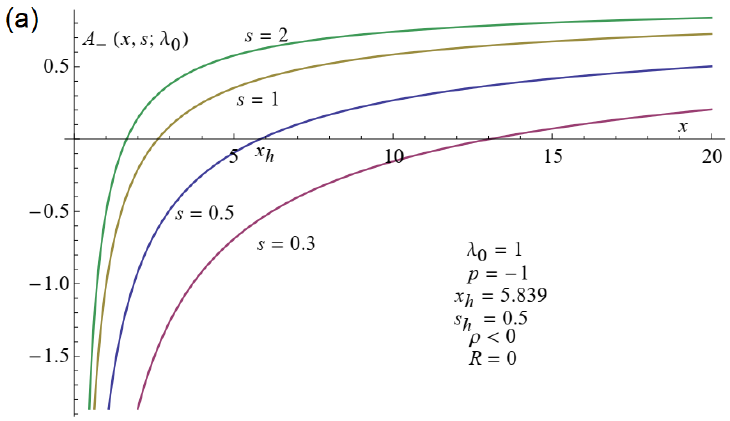} \includegraphics[scale=0.35]{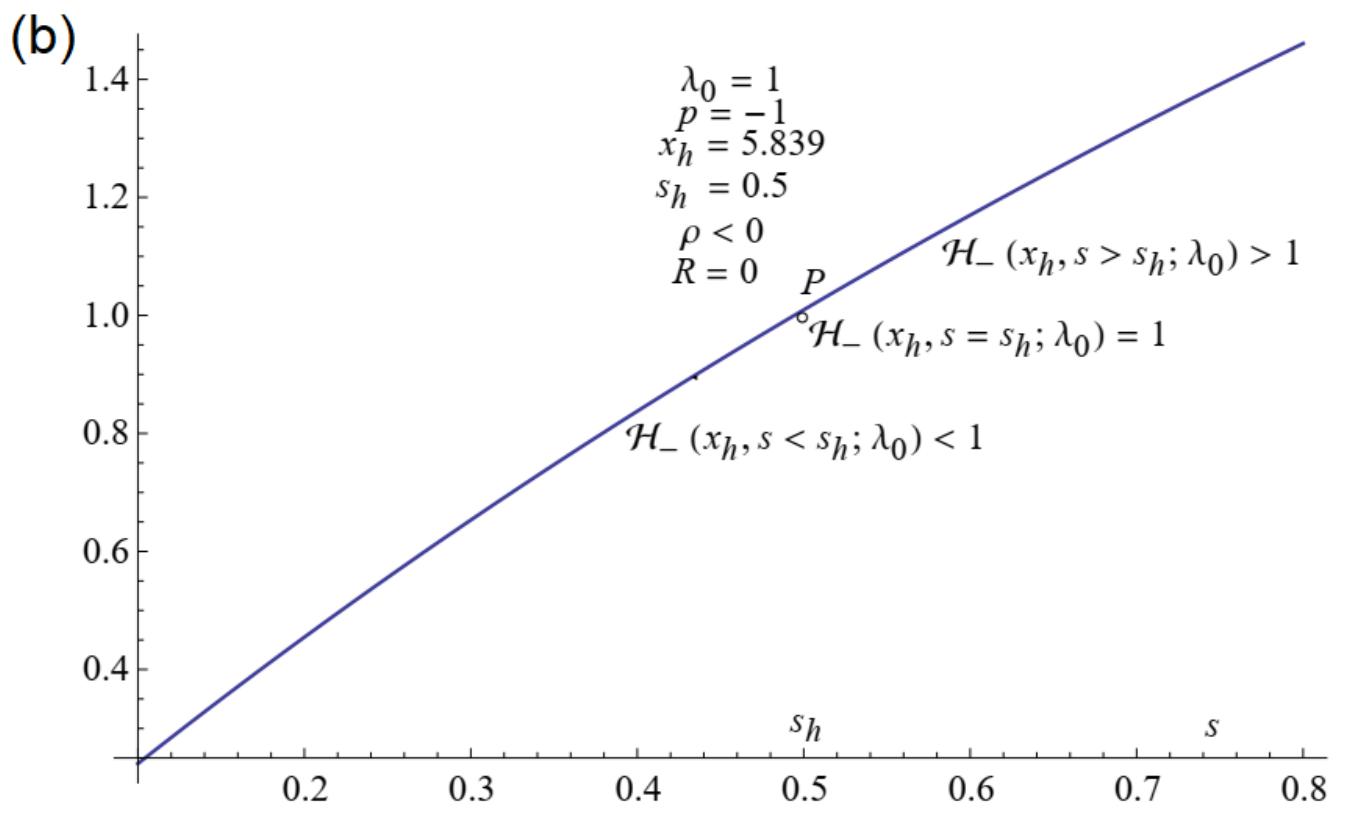}
\caption{(a): The chosen values in the metric function (21) are $\lambda _{0}=1$, $p=-1$ giving a single horizon tidal charge-type metric. Fig.7(b) shows once again that the hoop conjecture holds since $\mathcal{H}_{-}(x_{h},s\leq s_{h};\lambda _{0})\leq 1$ for any horizon radius $x_{h}$ corresponding to a choice $s_{h}$. The point $P$ bifurcates the spacetime into horizon and no-horizon regimes such that in the latter $\mathcal{H}_{-}(x_{h},s>s_{h}; \lambda _{0})>1$ holds.}\label{fig7}
\end{figure}

\begin{figure}[!ht]
\centering
\includegraphics[scale=0.70]{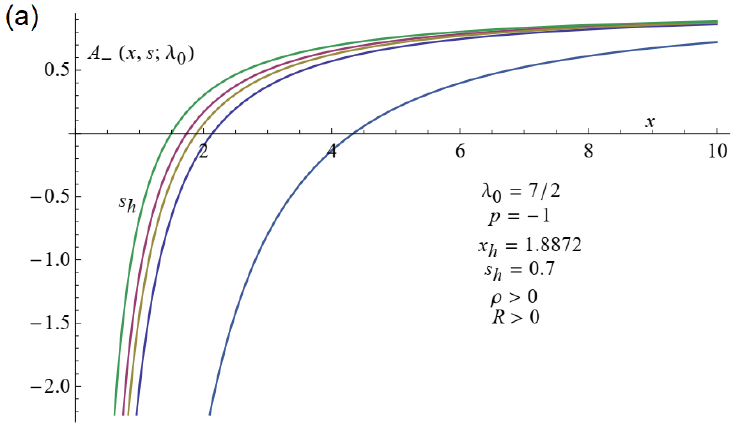} \includegraphics[scale=0.70]{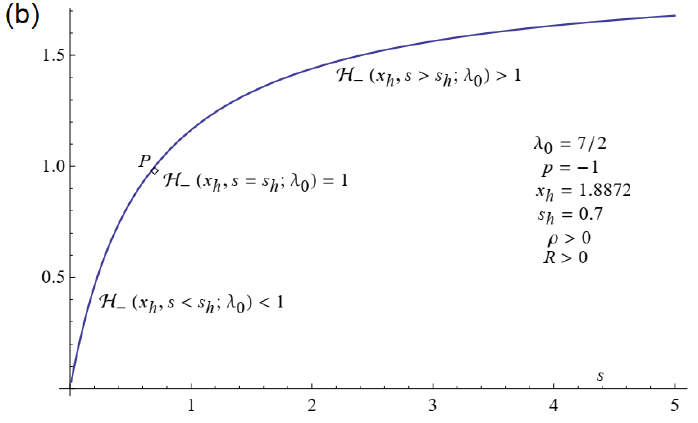}
\caption{(a): The chosen values in the metric function (21) are $\lambda _{0}=7/2$, $p=-1$ giving a single horizon generalized tidal charge-type metric. Fig.8(b) shows once again that the hoop conjecture holds since $\mathcal{H}_{-}(x_{h},s\leq s_{h};\lambda _{0})\leq 1$ for any horizon radius $x_{h}$ corresponding to a choice $s_{h}$. The point $P$ bifurcates the spacetime into horizon and no-horizon regimes such that in the latter $\mathcal{H}_{-}(x_{h},s>s_{h}; \lambda _{0})>1$ holds. Any other value $\lambda _{0}>0$ could be chosen at will and it can be verified that the horizon patterns and the hoop conjecture continue to be preserved.}\label{fig8}
\end{figure}

\section{Results}\label{sec5}
With the above algorithm at hand, we then plot the metrics, $A_{+}$ and $A_{-}$ for any fixed $\lambda _{0}$, and look for real critical values, if they exist. Then we shall graphically find the horizons, whether double or single, and plot the Hod functions to see if they satisfy the hoop conjecture $\mathcal{H}_{p}\leq 1$. Once again, we recall that $\Upsilon >0$ throughout the paper. Note that it is not possible to find analytic roots $x=x_{h}(s)$ for just any value of $\lambda _{0}$ by solving $A(x,s_{h};\lambda _{0})=0$ since roots beyond quartic algebraic equations are to be found only numerically, which is what we did in this paper.

Our results about the LSMA solutions are as follows: There appears four generic types of horizons for arbitrary values of $\lambda _{0}$ and $\Upsilon >0$ with the corresponding Hod functions developed in Eq.(19):

(A) Reissner-Nordstr\"{o}m type: This type shows the existence of real critical values ($x_{c},s_{c}$) and for the condition $s\leq s_{c}$, there appears double horizons on either side of the coincident horizon at $x=x_{c}$. There is a "no horizon regime" for $s>s_{c}$ as indicated by the conditions (26)--(28) from the plots of $A_{+}(x,s;\lambda _{0})$ vs $x$. The examples are asymptotically flat Reissner-Nordstr\"{o}m type black holes ($\lambda _{0}=1,p=+1)$ for which $\rho >0$, $\mathbf{R}=0$ [Fig.(1a)] or its generalization ($\lambda _{0}=2/3,p=+1)$ for which $\rho >0$, $\mathbf{R}<0$ [Fig.(2b)]. We verified that the conjecture holds for both choices of $\lambda _{0}$ in accordance with (i) of Sec.4, i.e., $\mathcal{H}_{+}(x_{c},s\leq s_{c};\lambda _{0})\mathcal{\leq }$ $1$ and $\mathcal{H}_{+}(x_{c},s>s_{c};\lambda _{0})\mathcal{>}$ $1$ occurs in the no horizon regime as indicated by (26)--(28) [Figs.(1b,2b)]. This was the method adopted also in Refs.\cite{Nandi:2020, Nandi:2022}, where the behavior of $\mathcal{H}_{+}$ depending on ranges of $s$ described the status of the conjecture.

(B) SdS type: This is an execeptional type, where real critical values ($x_{c},s_{c}$) do exist, but the inequalities on $s$ resulting from the plots of $A_{-}(x,s;\lambda _{0})$ vs $x$ for the occurrence of horizon regimes are exactly \textit{reverse} to the conditions (29)--(31). The examples are the SdS type black holes ($\lambda _{0}=-1,p=-1)$ for which $\rho >0$, $\mathbf{R}>0$ [Fig.3a] or its generalization ($\lambda _{0}=-4,p=-1)$ for which again $\rho >0$, $\mathbf{R}>0$ [Fig.4a]. There appear double horizons but it follows from Figs.(3b,4b) that the mass-circumference ratio is $\mathcal{H}_{-}(x_{c},s\geq s_{c};\lambda _{0})\mathcal{\geq }$ $1$ in complete violation of (ii) of Sec.4, \textit{thereby violating the hoop conjecture} with $\mathcal{H}_{-}$ given by Eq.(24).

(C) SAdS type: In this type, the critical values, instead of being real, become negative and/or imaginary, which are to be ruled out as unphysical by definition, $x>0$ and $s>0$. There is then no restriction on $s$ and it turns out that only single horizons appear corresponding to values of $s$. Thus, we arbitrarily choose any value of $s$, say, $s_{h}$, and from $A_{+}(x,s_{h};\lambda _{0})=0$, find numerically the corresponding root $x=x_{h}$. Examples are SAdS type ($\lambda _{0}=-1,p=+1)$ for which $\rho <0$, $\mathbf{R}<0$ [Fig.5a] or its generalization ($\lambda _{0}=-5,p=+1)$ for which $\rho <0$, $\mathbf{R}<0$ [Fig.6a]. We then verify that the conjecture holds in accordance with (iii) of Sec.4, i.e., $\mathcal{H}_{+}(x_{h},s\leq s_{h};\lambda _{0})\mathcal{\leq }$ $1$. Likewise, $\mathcal{H}_{+}(x_{h},s>s_{h};\lambda _{0})\mathcal{>}$ $1$ indicates no horizon regime [Figs.(5b,6b)].

(D) Braneworld type: The metric formally resembles asymptotically flat Reissner-Nordstr\"{o}m spacetime but with a negative charge \cite{Dadhich:2000}. We should clarify that this negative charge is not the electric charge of the usual Reissner-Nordstr\"{o}m spacetime but a tidal \textit{imprint} onto the $3d$-brane of the $5d$-bulk mediated by the conformal bulk Weyl tensor according to the Randall-Sundrum string model \cite{Randall:1999}. This imprint is called "tidal charge" and its role is to introduce corrections to the Schwarzschild potential, similar to the corrections obtained by LSMA \cite{Lessa:2020} in the Kalb-Ramond model \cite{Kalb:1973}. However, there is a major difference as to how the two corrections alter the horizon radii. While in the Kalb-Ramond model, one has $p=+1$ yielding double horizons [type (A) above]; in the tidal case, on the other hand, one has $p=-1$ yielding only single horizons of $A_{-}(x,s;\lambda _{0})$ corresponding to the chosen values of $s$, say $s_{h}$. The braneworld tidal charge black hole has ($\lambda_{0}=1,p=-1)$ for which $\rho <0$, $\mathbf{R}=0$ [Fig.7a] or its generalization ($\lambda _{0}=7/2,p=-1)$ for which $\rho >0$, $\mathbf{R}>0$ [Fig.8a]. We verify that the conjecture holds in accordance with (iii) of Sec.4, i.e., $\mathcal{H}_{-}(x_{h},s\leq s_{h};\lambda _{0})\mathcal{\leq}$ $1$. Likewise, $\mathcal{H}_{-}(x_{h},s>s_{h};\lambda _{0})\mathcal{>}$ $1 $ indicates no horizon regime [Figs.(7b,8b)].

In all the above four cases, the point $P$ on the plots of the Hod function [Figs.1b -8b] display the point of transition between horizon and no horizon regimes thus radially bifurcating the spacetime into two regions across the point $P$. Also, while the signs of $\lambda $ and $p$ are preserved in each category, the behavior of the source stresses and the Ricci scalar $\mathbf{R}$ can differ drastically, but the number of horizons (double or single) and the behavior of the Hod function remain the same. Values of $\lambda_{0}$ have been changed at will in the displayed plots but any other different value could as well be chosen with the resulting qualitative features falling into one of the above four mutually exclusive categories determined by the combinations of $\lambda =1,-1$ and $p=+1,-1$. So long as signs are maintained, different numerical values of $\lambda $ are allowed maintaining the number of ($\lambda ,p$) combinations to only $2^{2}=4$.

The analytic expressions for horizon radii in the simplest Reissner-Nordstr\"{o}m type ($\lambda _{0}=1,p=+1)$, SdS-type ($\lambda _{0}=-1,p=-1$), SAdS-type ($\lambda _{0}=-1,p=+1)$ and Braneworld-type ($\lambda _{0}=1,p=-1)$ black holes can be found in the literature and needless to say that the corresponding profiles indicated in the Figs.(1a-8a) conform to those analytic expressions. For generalized versions, however, one has to find the horizons only numerically.

\section{Summary and remarks}\label{sec6}
The enlarged class of LSMA black holes, studied here numerically, offers interesting examples of black holes with single or double horizons for \textit{arbitrary} values of LV parameters - only a few examples are given here but the conclusions are quite generic as can be verified on a case by case basis since analytic roots giving horizon radii are not available beyond quartic algebraic equations. The status of Thorne's hoop conjecture illuminates the nature of concerned spacetimes, in particular, the SdS black hole, in which horizons do exist but the conjecture does not hold. The details are as follows.

If a static spherically symmetric metric has only one constant parameter, say, $M_{\infty }$ with a single horizon, such as the Schwarzschild black hole, then the hoop conjecture holds in its extreme form with the circumference-mass ratio on the horizon being given by $\mathcal{H=}\frac{R}{2M_{\infty }}=1$. But if the single horizon spacetime has, apart from $M_{\infty}$, another independent parameter $\Upsilon $, then the ratio $\mathcal{H}$ can no longer hold in its extreme form but should include also the values of $\Upsilon $ since the horizon radii are shifted by it, as analytically exemplified by LSMA for the case $\lambda =1$ \cite{Lessa:2020}. In the present work, the two parameters, $M_{\infty }$ and $\Upsilon $, were equivalently described by two dimensionless parameters ($x,s$) in the metric (21) for any arbitrarily chosen $\lambda $. The ($x,s$) enter into the quasi-local mass (6) used in defining the conjecture. Then, for \textit{every} horizon radius at $x=x_{h}$, there will be a range of admissible values of $s$ such that the conjecture would be satisfied, i.e., $\mathcal{H} (x_{h},s\leq s_{h})\mathcal{\leq }$ $1$ would hold according to (iii) of Sec.4. The extreme equality holding only at $s=s_{h}.$The same arguments apply also to double horizon spacetimes as stated in (i) and (ii) of Sec.4 except that there is now $\left( x_{c},s_{c}\right) $ instead of $\left(x_{h},s_{h}\right) $.

The LSMA solutions represent LV corrected black holes with Schwarzschild mass $M_{\infty }$ for arbitrary values of the LV parameters $\lambda$ and $\Upsilon$. Combinations of arbitrary values of LV parameters may as well lead to exotic source matter (i.e., $\rho <0$) and there is no \textit{a priori} reason to rule out such black holes. Some well known basic examples of this kind are the SAdS black hole having wormhole-like topology ($\rho <0, \mathbf{R}<0$) and the \textit{braneworld }"tidal charge" black hole ($\rho<0$, $\mathbf{R}=0$), both formally belonging to the LSMA class of solutions for $\lambda =-1$ and $1$ respectively (Figs.3,4). The example for $\lambda =1$ (Figs.4) is particularly intriguing since the LV parameter $\Upsilon >0$ can now be interpreted to play the role of a tidal charge, an imprint from the $5d$ bulk onto the $3d$ brane in the Randall-Sundrum scenario \cite{Randall:1999}.

While Schwarzschild horizon itself satisfies the extreme form of the conjecture, it is not obvious if its LV corrected version would satisfy the conjecture for arbitrary values of $\lambda $ and $\Upsilon$. To investigate this question, we first graphically found the horizon patterns (i.e., coincident/double or single) for some arbitrary values of LV parameters. The horizons have been classified into \textit{four mutually exclusive generic} types (A,B,C,D) based on the combinations of $\lambda =1,-1$ and $p=+1,-1$. In addition to these values, some other illustrative numerical values of $\lambda $ were also chosen and the metric functions plotted as indicated in Figs.(1a-8a). Next, we verified the status of the conjecture for these spacetimes by plotting the relevant Hod function $\mathcal{H}$ developed in the generic format for arbitrary ($x,s$). It has been found that, while the hoop conjecture $\mathcal{H}$ $\leq 1$ holds for three types (A,C,D), it is \textit{violated} in the spacetime type B, i.e., in the SDS type black holes (Figs.3b,4b).

We point out that the braneworld tidal charge type of LSMA black hole \textit{increases} the LV correction to planetary perihelion advance in contrast to the decrease due to ordinary black holes ($\lambda =1,p=+1,\rho > 0$) providing a qualitative distinction between them. The general relativistic leading order perihelion precession of planets in the Schwarzschild field is known as $\delta \Phi _{GR}=\frac{6\pi M_{\infty }}{\ell }$ (where $\ell $ is the semi-latus rectum of the Keplerian orbit) and the LV correction calculated in \cite{Lessa:2020} is given by $\Delta \Phi _{LV}^{\lambda =1}=-\frac{2\pi p\Upsilon }{M_{\infty }\ell }<0$, which for $p=+1$ shows that the LV correction decreases $\delta \Phi _{GR}$. On the other hand, for tidal charge braneworld black holes ($\lambda =1$, $p=-1,\rho <0$), one has the LV correction $\Delta \Phi _{p=-1}^{\lambda =1}=\frac{2\pi \Upsilon }{M_{\infty }\ell }>0$ implying an increase in the GR precession value. This phenomenon qualitatively distinguishes non-exotic from "exoticized" black holes. For the latter black hole of a given Schwarzschild mass $M_{\infty }$, the physical condition $\delta \Phi _{GR}>>\Delta \Phi _{p=-1}^{\lambda=1} $ yields a constraint on the shift of the horizon for a black hole of mass, say, $10$ times more massive than the solar mass $M_{\infty }$ given by $\frac{\Upsilon }{\left( 2\times 10M_{\infty }\right) ^{2}}<<7.5\times 10^{-3} $. This constraint is\textit{\ independent} of the orbit size and is consistent with but weaker than the constraint $\frac{\Upsilon }{\left(2\times 10M_{\infty }\right) ^{2}}$ $\approx 2\times 10^{-6}$ obtained in \cite{Lessa:2020} from the comparison of planetary precession data.

Finally, we make a couple of speculative remarks: First, it is indeed thought provoking that Kalb-Ramond model \cite{Lessa:2020, Kalb:1973}, Randall-Sundrum braneworld scenario \cite{Randall:1999, Dadhich:2000} or even massive gravity \cite{Bebronne:2009, Jusufi:2018} yield \textit{formally} similar power law modifications although the parent theories are ideologically very different. It should be quite rewarding to examine in the future whether or not there could be some deeper connection between these fundamental theories or the similtude of the solutions is merely a PPN type power law coincidence. Second, we found that the hoop conjecture is violated in the double horizon asymptotically non-flat SdS type black holes or in its generalized versions. It is quite tempting to speculate that \textit{this violation could be a generic feature of any multi-horizon asymptotically non-flat spacetimes} (see Figs 3b,4b). This violation seems consistent with the problematic thermodynamic picture in the SdS spacetime discussed in detail in \cite{Roy:2007}.

\section*{Acknowledgments}

We are indebted to two anonymous referees for their helpful comments. This work was supported by the Russian Science Foundation under grant no. 23-22-00391, https://rscf.ru/en/project/23-22-00391/.

\end{document}